\documentclass[prb,superscriptaddress,twocolumn]{revtex4-2}
\usepackage[utf8]{inputenc}
\usepackage[T1]{fontenc}

\usepackage{amsmath}
\usepackage{amssymb}
\usepackage{newtxtext}
\usepackage[smallerops]{newtxmath}
\let\lambda\lambdaup
\allowdisplaybreaks
\usepackage{graphicx}
\graphicspath{{fig/}}
\usepackage{xcolor}
\usepackage[colorlinks=true,linkcolor=blue,citecolor=blue,urlcolor=blue]{hyperref}
\usepackage[capitalise]{cleveref}
\usepackage{multirow}

\renewcommand{\Re}{\mathop{\mathrm{Re}}}

\newcommand{\var}{\mathop{\mathrm{var}}}
\newcommand{\cov}{\mathop{\mathrm{cov}}}
\newcommand{\relu}{\mathop{\mathrm{relu}}}
\newcommand{\ph}{\phantom0}
\newcommand{\phs}{\ph\ph}

\begin{document}

\title{High-accuracy variational Monte Carlo for frustrated magnets with deep neural networks}

\author{Christoper Roth}
\address{Physics Department, University of Texas at Austin}
\author{Attila Szab\'{o}}
\address{Rudolf Peierls Centre for Theoretical Physics, University of Oxford, Oxford OX1 3PU, UK}
\address{ISIS Facility, Rutherford Appleton Laboratory, Harwell Campus, Didcot OX11 0QX, UK}
\author{Allan H. MacDonald}
\address{Physics Department, University of Texas at Austin}

\date{\today}

\begin{abstract}
We show that neural quantum states based on very deep (4--16-layered) neural networks can outperform state-of-the-art variational approaches on highly frustrated quantum magnets, including quantum-spin-liquid candidates.
We focus on group convolutional neural networks (GCNNs) that allow us to efficiently impose space-group symmetries on our ansätze.
We achieve state-of-the-art ground-state energies for the $J_1-J_2$ Heisenberg models on the square and triangular lattices, in both ordered and spin-liquid phases, and discuss ways to access low-lying excited states in nontrivial symmetry sectors. 
We also compute spin and dimer correlation functions for the quantum paramagnetic phase on the triangular lattice, which do not indicate either conventional or valence-bond ordering.      
\end{abstract}

\maketitle

\section{Introduction}

Frustrated magnetism has long been an exciting playground for discovering new physics.
Perhaps the most interesting outcome of this activity has been the concept of a \textit{quantum spin liquid}, a phase of matter without any long-range order even in the ground state~\cite{Balents2010SpinMagnets,Savary2017QuantumReview},
characterised by long-range entanglement, fractionalised excitations, and topological order. 
Establishing spin-liquid physics in specific microscopic models, however, remains a great theoretical challenge, apart from a few exactly solvable systems and perturbative constructions~\cite{Kitaev2003Fault-tolerantAnyons,Kitaev2006AnyonsBeyond,Hermele2004PyrochloreMagnet}:
While spin-liquid phases have been proposed in Heisenberg and Hubbard models on a number of lattices~\cite{Hu2013DirectAntiferromagnetism,Nomura2021Dirac-TypeSpectroscopy,Liu2022SquareGaplessQSL,Liu2022SquareDQCP,Iqbal2016TriangularJ1J2,Hu2019DiracAntiferromagnet,Wietek2017TriangularHeisenbergCSL,Yan2011Spin-liquidAntiferromagnet.,Depenbrock2012NatureLattice,Iqbal2013GaplessAntiferromagnet,He2017SignaturesModel,Lauchli2019TheRevisited,Kiese2022KagomeSkymap,Szasz2020TriangularHubbard,mezzacapo2012ground}, their presence and properties remain a matter of active debate.
To a large extent, this is due to the difficulty of reliable numerical simulations of frustrated magnets:
unbiased quantum Monte Carlo is plagued by the sign problem~\cite{Troyer2005ComputationalSimulations},
while variational approaches (including tensor networks) are susceptible to bias towards particular kinds of state that may override the often minuscule energy differences~\cite{Lauchli2019TheRevisited} between competing spin-liquid and  ordered ground states.

An interesting approach to overcome this difficulty is employing neural networks to parametrise variational wave functions~\cite{Carleo2017SolvingNetworks}.
Neural networks are universal function approximators~\cite{hornik1989multilayer}, thus they can represent arbitrarily complex quantum many-body wave functions, without an obvious bias for particular kinds of state.
Furthermore, variational computations using such neural quantum states (NQS) benefit from the greater computational power of graphics processing units (GPUs) and powerful machine-learning optimisation libraries~\cite{Vicentini2022NetKetSystems}.
NQS ansätze based on a variety of neural-network architectures have been proposed~\cite{Choo2019StudyStates,Szabo2020NeuralProblem,Hibat-Allah2020RecurrentWavefunctions,Sharir2020DeepSystems,Astrakhantsev2021Pyrochlore,Roth2021GroupAccuracy},
however, they generally fall short of the reliability and accuracy required for state-of-the-art results on frustrated problems.
A remarkable exception is the ansatz developed in Refs.~\cite{Nomura2021Dirac-TypeSpectroscopy,Nomura2017GutzwillerRBM}, which, however, uses neural networks merely to improve the accuracy of Gutzwiller-projected fermionic wave functions, an extremely successful ansatz in its own right.

By contrast, we demonstrate here that a ``pure'' NQS ansatz using very deep networks can achieve state-of-the-art variational energies.
In particular, we use \textit{group-convolutional neural networks (GCNNs)}~\cite{Cohen2016GroupNetworks,Roth2021GroupAccuracy}, which allow us to impose the full space-group symmetry of the problem on the wave functions.
We find two key design principles of a successful architecture:
First, the network should not separate the amplitudes and phases of the network, as learning the latter in a frustrated system is beyond the capacity of even deep neural networks~\cite{Westerhout2020,Bukov2021Learning}.
Second, imposing locality by using short-ranged convolutional filters in the GCNNs both makes using deeper networks computationally feasible and simplifies the learning landscape by structuring the representation of long-range correlation in the networks; the latter is reflected in faster convergence compared to full-width convolutional kernels.
Our ansätze either match or surpass the best variational energies in the literature~\cite{Nomura2021Dirac-TypeSpectroscopy,Iqbal2016TriangularJ1J2} in the quantum paramagnetic regimes of the square- and triangular-lattice $J_1-J_2$ Heisenberg model on clusters as large as $16\times16$.
We also present numerical experiments on finding low-lying excited states using GCNNs.

\section{Group-convolutional neural networks} 

Lattice Hamiltonians are invariant under a large group of spatial symmetries, governed by the geometry of the lattice and anisotropies of the interactions:
Wigner's theorem~\cite{wigner2012group} ensures that all eigenstates of such a Hamiltonian transform under an irreducible represenation (irrep) of the same symmetry group.
Imposing space-group symmetries explicitly in a variational ansatz reduces the Hilbert space available for the variational algorithm, making it more reliable and efficient~\cite{dAscoli2019FindingBias}.
Furthermore, symmetry-broken phases are identified by symmetry quantum numbers of the lowest-lying excited states, known as the tower of states~\cite{Anderson1952TOS,Wietek2017TOS}, thus access to the lowest-lying states in a number of symmetry sectors  allow identifying distinct phases and transition points with a high accuracy~\cite{Nomura2021Dirac-TypeSpectroscopy}.

Space-group symmetries can be imposed on variational wave functions by explicit projection.
Consider an ansatz $\psi_0(\vec\sigma;\vec\theta)$ with parameters $\vec\theta$ that maps a computational basis state $\vec\sigma$ onto its amplitude in the variational wave function. 
Then, given a group $G$ of symmetries $\hat g$ that maps each basis state $\vec \sigma$ onto another, $\hat g\vec\sigma$, we can construct an ansatz transforming under a given irrep of $G$ using the projection formula~\cite{Heine1960GroupMechanics}
\begin{subequations}
\begin{align}
    |\psi\rangle &= \frac{d_\chi}{|G|}\sum_{g\in G} \chi_g^* \hat g|\psi_0\rangle\\
    \psi(\vec\sigma) &= \frac{d_\chi}{|G|}\sum_{g\in G} \chi_g^* \psi_0(\hat g^{-1}\vec\sigma),
    \label{eq: irrep projection basis states}
\end{align}
\label{eq: irrep projection formula}%
\end{subequations}
where $\chi_g$ are the characters of the irrep. 

Projection approaches based on~\eqref{eq: irrep projection formula} have been used with a variety of ansätze~\cite{Nomura2021Dirac-TypeSpectroscopy,Choo2019StudyStates,Misawa2019MVMCOpen-sourceMethod}.
Usually, however, one has to evaluate $\psi_0$ for all symmetry-related basis states $\hat g\vec\sigma$:
this can be prohibitively expensive, requiring setting all or most $\psi_0(\hat g\vec\sigma)$ equal by construction, which in turn restricts which irreps of the symmetry group can be probed with the ansatz.
Instead, one would prefer to generate all $\psi_g(\sigma) \equiv \psi_0(\hat g^{-1}\vec\sigma)$ in a single evaluation of the ansatz.
Such an ansatz would be \textit{equivariant} under the symmetry group $G$ in the sense that acting on its input by a symmetry element would cause the output to be acted on by the same symmetry in a nontrivial way (namely, by the left canonical action):
\begin{equation}
    \psi_g(\hat h^{-1}\vec\sigma) = \psi_0(\hat g^{-1}\hat h^{-1} \vec\sigma) = \psi_{hg}(\vec\sigma).
    \label{eq: equivariance}
\end{equation}
Note that $(\hat h^{-1}\vec\sigma)_{\vec r} = \sigma_{\hat h\vec r}$, so such an equivariant ansatz maps relabelling lattice sites by $\hat h$ to left-multiplication of symmetry elements by $h$.

Convolutional neural networks (CNNs) are a famous example of such an equivariant function. Let us consider a (hypercubic) lattice of size $L_1\times L_2\times \dots \times L_d$ in periodic boundary conditions and its translation group, $\mathbb{Z}_{L_1}\times\dots\times\mathbb{Z}_{L_d}$, which has the same structure as the lattice itself. Now, the convolutional mapping
\begin{equation}
    \psi(\vec r) = \sum_{\vec r'} K(\vec r-\vec r') \sigma(\vec r'),
    \label{eq: convolution}
\end{equation}
with an arbitrary kernel $K$, is manifestly equivariant:
\begin{equation*}
    \sum_{\vec r'} K(\vec r-\vec r') \sigma(\vec r' + \vec h) = \sum_{\vec r'} K(\vec r + \vec h - \vec r') \sigma(\vec r') = \psi(\vec r+\vec h),
\end{equation*}
where vector addition winds around the periodic boundary conditions. This is equivalent to~\eqref{eq: equivariance}, as a translation $\hat h$ acts by adding $\vec h$ to both input and output coordinates.
One can use similar arguments to show that several iterations of the convolution~\eqref{eq: convolution}, arbitrary ``on-site'' functions $y(\vec r) = f[x(\vec r)]$, and thus arbitrarily deep neural networks built from alternating these building blocks, are all equivariant.
The projection formula~\eqref{eq: irrep projection formula} also has a natural interpretation. 
The irreps of the translation group are all one-dimensional and labelled by the phases $e^{i\phi_1}, \dots, e^{i\phi_d}$ acquired by the wave function upon translation along each lattice vector: \cref{eq: irrep projection formula} thus becomes
\begin{equation*}
    \psi(\vec\sigma) = \frac1{L_1\dots L_d} \sum_{\vec r} e^{-i\vec \phi\cdot\vec r} \psi_{\vec r}(\vec \sigma),
\end{equation*}
a Fourier transform that extracts a crystal momentum eigenstate.

\textit{Group-convolutional neural networks (GCNNs)}~\cite{Cohen2016GroupNetworks,Roth2021GroupAccuracy} generalise this idea to arbitrary (nonabelian) groups.
Since the group is no longer isomorphic to the input lattice, we will require two types of linear layer:
first, an embedding layer
\begin{subequations}
\begin{equation}
    y_g = \sum_{\vec r} K(\hat g^{-1} \vec r) \sigma(\vec r)
    \label{eq: embedding layer}
\end{equation}
generates equivariant feature maps, indexed with group elements, from the input; 
this is followed by any number of group convolutional~\footnote{Our convention differs from that of Ref.~\cite{Cohen2016GroupNetworks}, which in fact implements group correlation rather than convolution. The two conventions are equivalent (the indexing of the kernels differs by taking the inverse of each element).} layers
\begin{equation}
    z_g = (W\circ y)_h \equiv \sum_{h\in G} W(h^{-1}g) y_h
    \label{eq: GC layer}
\end{equation}
and ``on-site'' nonlinearities. One can show~\cite{Vicentini2022NetKetSystems} that such a network satisfies~\eqref{eq: equivariance}. 
\end{subequations}

A naïve alternative to GCNNs for space groups would be applying all point-group operations (rotations, reflections, etc.) on $\vec\sigma$ and feeding each of these configurations into a conventional CNN~\cite{Choo2019StudyStates}.
This too satisfies~\eqref{eq: equivariance} and can be projected onto any irrep using~\eqref{eq: irrep projection formula}: In fact, it is equivalent to a GCNN with the same number of features in each layer, in which every kernel $W$ is restricted to pure translations rather than the whole space group. Thus, GCNNs of the same size can have more variational parameters, and in every layer, each output is determined by more inputs from the previous layers: both of these features allow the network to be more expressive at the same memory footprint. Indeed, GCNNs have outperformed symmetrised CNNs in both image-recognition~\cite{Cohen2016GroupNetworks} and variational Monte Carlo~\cite{Roth2021GroupAccuracy} tasks.

\subsection{Details of the GCNN ansatz}
\label{sec: architecture}

\begin{figure}
    \centering
    \includegraphics{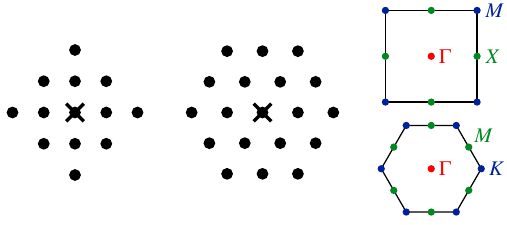}
    \caption{Arrangement of nonzero kernel entries for the square (left) and triangular (centre) lattices. The cross indicates the origin of the lattice and/or pure point-group symmetries.
    Right: illustration of high-symmetry points in the Brillouin zones of the square (top) and triangular (bottom) lattices.}
    \label{fig: kernels}
\end{figure}

In this paper, we will consider spins on a lattice and represent their wave functions in the $\sigma^z$ basis. 
The largest symmetry group for which~\eqref{eq: irrep projection basis states} is applicable in this basis is the direct product of the space group of the lattice and the $\mathbb{Z}_2$ spin parity group generated by $P=\prod_i \sigma_i^x$.%
\footnote{The full $SU(2)$ spin-rotation symmetry of Heisenberg models can only be imposed for a few ansätze that show $SU(2)$ invariance explicitly~\cite{Vieijra2020RestrictedSymmetries,vieijra2021many,Misawa2019MVMCOpen-sourceMethod}. We note that for $M_z=0$, states whose total spin quantum number $S$ is even transform as $P=+1$, while odd-$S$ states have $P=-1$.}.
Irreps of this group are specified by the eigenvalue of $P=\pm1$ and an irrep of the space group.
The latter are generally specified by a set of crystal momenta related by point-group operations (known as a \textit{star}) and an irrep of the subgroup of the point group that leaves these momenta invariant (known as the \textit{little group})~\cite{bradleycracknell1972,internationaltablesA,Vicentini2022NetKetSystems}. 
We shall denote space-group irreps using a representative wave vector of the star and the Mulliken symbol~\cite{Mulliken1933} of the little-group irrep: for example, the trivial irrep of the square-lattice space group is written as $\Gamma.A_1$ (cf.~Fig.~\ref{fig: kernels}).
Parity eigenvalues will be included through a $\pm$ index, e.g., $\Gamma.A_1^+$

All GCNN ansätze used in this work have a fixed number of feature maps in every layer, connected by real-valued kernels $K,W$.
In most cases, we restrict the embedding kernels $K$ to the local clusters shown in \cref{fig: kernels}.
We also impose locality on the convolutional kernels $W$: these are only nonzero for $g=tp$, where $p$ is a point-group symmetry that leaves the origin in place, and $t$ is a translation by a lattice vector indicated in \cref{fig: kernels}.

All but the last layer is followed by the \textsc{selu} activation function~\cite{selu2017}, which allows us to train very deep (up to 16 layers in this work) networks without severe instabilities.
In the output layer, we combine pairs of feature maps into complex-valued features, exponentiate them, and project the result on the desired irrep:
\begin{subequations}
\begin{align}
    \tilde h_{n,g} &= h_{n,g} + ih_{n+F/2,g} \\
    \psi(\vec\sigma) &= \sum_{g\in G} \chi_g^* \sum_{n=1}^{F/2} \exp(\tilde h_{n,g}),
\end{align}
\label{eq: detailed ansatz}%
\end{subequations}
where $F$ is the number of (real-valued) feature maps.
Including exponentiation in~\eqref{eq: detailed ansatz} is important to represent the wide dynamical range of wave-function amplitudes: for example, the amplitude of highly ferromagnetically correlated basis states is exponentially suppressed in the ground state of antiferromagnets~\cite{Huse1988Simple}, which neural networks do not represent efficiently without any structural bias~\cite{Trask2018NeuralUnits}. 
Complex exponentiation, however, separates amplitudes and phases, which obstructs the learning of the accurate sign structure, a formidably complex object for a highly frustrated magnet~\cite{Westerhout2020,Szabo2020NeuralProblem,Valle-Perez2018DeepFunctions}:
constructing $\psi$ as a sum of many terms alleviates this problem, as cancellation between the terms allows learning destructive interferences more easily~\cite{Nomura2021HelpingRBM}.

\subsection{GCNNs with residual layers}
\label{sec: residual gcnn}

\begin{figure}
    \centering
    \includegraphics{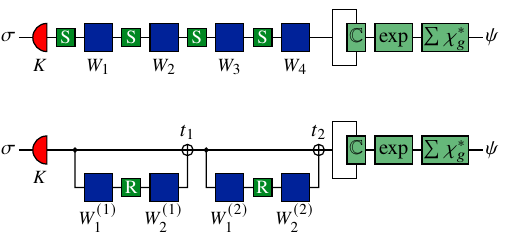}
    \caption{Structure of the plain (top) and residual (bottom) GCNNs used in this work. Red and blue boxes stand for the embedding layer~\eqref{eq: embedding layer} and the group convolutions~\eqref{eq: GC layer}, respectively. Dark green boxes indicate \textsc{selu} (S) or \textsc{relu} (R) activation functions. Light green boxes represent the output layer~\eqref{eq: detailed ansatz}. }
    \label{fig: networks}
\end{figure}

While the GCNN wave function described above performs quite well on the square lattice, we found that it struggles to converge to state-of-the-art energies on the more challenging triangular-lattice models.
We remedied this problem using residual group-convolutional layers: 
These make deep networks more expressive, and their training more stable, by allowing the network to effectively remove superfluous layers~\cite{he2016deep}.
The hidden layers of our residual GCNN (ResGCNN) are given by
\begin{equation}
    \mathbf{h}^{(i+1)} = \mathbf{h}^{(i)} + t_i W_2^{(i)} \circ \relu\left[ W_1^{(i)} \circ \mathbf{h}^{(i)}\right];
\end{equation} 
the embedding layer~\eqref{eq: embedding layer} and the construction~\eqref{eq: detailed ansatz} of $\psi$ from the last hidden layer remains unchanged.
The structures of plain and residual GCNNs are compared in Fig.~\ref{fig: networks}.
The trainable variables $t_i$ allow the training protocol to easily control the relative importance of each layer. We initialise $t_i=0$, so that all hidden layers start out as trivial.

\subsection{Training protocol}
\label{sec: training protocol}

We optimise our variational wave function using the stochastic reconfiguration (SR) method~\cite{sorella1998green}, in which the parameter updates $\delta\theta$ are found by solving the equation
\begin{equation}
    S\, \delta\vec \theta = -\eta\frac{\partial E}{\partial \vec\theta},
    \label{eq: basic SR}
\end{equation}
where $E$ is the variational energy $\langle \psi | H|\psi\rangle/\langle \psi|\psi\rangle$, and $S$ is the quantum geometric tensor~\cite{Stokes2020quantumnatural}.
\cref{eq: basic SR} is ill-conditioned, requiring regularisation of the $S$ matrix~\cite{becca_sorella_2017}. 
We found that the standard approach of adding a constant to diagonal entries leads to poor convergence, while the scale-invariant regulariser of Ref.~\cite{becca_sorella_2017} is itself numerically unstable.
Adding both types of shift, on the other hand, stabilises calculations and allows reliable convergence. We thus use $S_{ii} \mapsto S_{ii} + \varepsilon_1 S_{ii} + \varepsilon_2$ and set $\eta=0.01, \varepsilon_1=10^{-2}, \varepsilon_2=10^{-3}$ in all simulations. 
We have not fine-tuned these hyperparameters and expect a wide range of $\eta$ and $\varepsilon_2\lesssim\varepsilon_1$ to yield similar results.

The above scheme works reliably relatively near the ground-state energy but it is often unstable at the start of training.
We found that this is greatly improved by biasing the training towards states with similar amplitudes for all basis states in this stage. 
A physically motivated approach is minimising a ``free energy'' $F=E-T\mathcal{S}$, where the ``entropy'' is defined in the computational basis as~\cite{roth2020iterative,hibat2022supplementing}
\begin{equation}
    \mathcal{S} = - \sum_{\vec\sigma} \frac{|\psi(\sigma)|^2}{\sum_{\vec\tau} |\psi(\tau)|^2} \log \frac{|\psi(\sigma)|^2}{\sum_{\vec\tau} |\psi(\tau)|^2};
    \label{eq: entropy}
\end{equation}
the denominators are necessary as our wave functions are generally not normalised.
Even though we cannot calculate $\mathcal{S}$ directly because of this, its gradients can be estimated via Monte Carlo sampling, see \cref{sec: gradients}.
The effective temperature $T$ is gradually lowered to zero as the training proceeds: we used $T_n = \exp(-n/50)/2$ in the $n$th step.

To find low-energy variational states in nontrivial symmetry sectors,
we used networks of the same architecture as in the ground-state simulations and initialised them with the converged parameters of the latter.
This amounts to using the ground state to generate an initial guess for the excited state by changing the irrep characters $\chi$ in~\eqref{eq: irrep projection formula}. 
We found that this transfer learning process starts at variational energies close to that of the ground state and converge quickly (100--300 SR steps with 4096 Monte Carlo samples) to a stable variational energy; we used 1000 steps to allow for full annealing of the energy measurement.
By contrast, networks trained from scratch either became unstable or levelled off at extremely high energies.

All simulations were carried out on a single NVIDIA A100 GPU.

\section{Square lattice $J_1-J_2$ model}

We first apply our approach to the square-lattice Heisenberg model with first- and second-neighbour couplings,
\begin{equation}
H = J_1 \sum_{\langle ij \rangle} \vec s_i \cdot \vec s_j + J_2 \sum_{\langle \langle ij \rangle \rangle} \vec s_i \cdot \vec s_j;
\end{equation}
we will set $J_1=1$.
The model orders both at small and large values of $J_2$, showing a Néel and a stripy pattern, respectively.
Near the classically maximally frustrated point, $J_2/J_1=0.5$, both of these orders disappear: the current consensus is that this region is split between a $\mathbb{Z}_2$ Dirac spin liquid and a valence-bond solid (VBS)~\cite{Nomura2021Dirac-TypeSpectroscopy,Liu2022SquareGaplessQSL,Shackleton2022SquareZ2}. 
We benchmark our method at $J_2/J_1=0.5$ and 0.55, thought to lie inside the spin-liquid and VBS phases, respectively.

\subsection{Ground-state energies}
\label{sec: square GS}

We estimated the ground state energy for the square-lattice $J_1-J_2$ model for linear cluster sizes $L=10,12,16$. 
We used GCNNs with $L$ layers, each consisting of six group-indexed feature maps.
For most simulations, we performed 1000 SR steps using 1024 Monte Carlo samples, followed by 500 steps with 4096 samples.
The first stage allows approaching the ground state at relatively low computational cost, while training with more samples helps the network learn representations of the wave function that generalise better to portions of the Hilbert space that are not sampled~\cite{Westerhout2020}.

\begin{table}
    \centering
    \setlength{\tabcolsep}{0.5em}
    \begin{tabular}{cclcc} \hline\hline
        $J_2/J_1$ & System size & \multicolumn{1}{c}{Our work} & Ref.~\cite{Nomura2021Dirac-TypeSpectroscopy} & $\langle \vec S^2\rangle$ \\\hline
        \multirow{3}{*}{0.5} & $10\times10$ & $-0.497437(7)$ & $-0.497629(1)$ & 0.038(4)\ph \\
         & $12\times12$ & $-0.496769(9)$ & $-0.496791(4)$ & 0.098(8)\ph \\
         & $16\times16$ & $-0.496509(6)$ & $-0.496213(3)$ & 0.156(10) \\\hline
        \multirow{3}{*}{0.55} & $10\times10$ & $-0.486772(11)$ & ---  & 0.077(7)\ph \\
         & $12\times12$ & $-0.486068(10)$ & $-0.485735(7)$ & 0.167(10) \\
         & $16\times16$ & $-0.485583(8)$ & $-0.485208(4)$ & 0.307(13) \\\hline\hline
    \end{tabular}
    \caption{Optimised ground-state energies (in units of $J_1/\textrm{spin}$) and estimates of the total spin $\langle S^2\rangle$ for the $J_1-J_2$ square-lattice Heisenberg model, compared to the best known variational energies for these systems~\cite{Nomura2021Dirac-TypeSpectroscopy}.}
    \label{tab: square GS}
\end{table}

Our best variational energies are summarised in Table~\ref{tab: square GS}. 
In all experiments, the converged variational energies come very close to the best ones currently available in the literature~\cite{Nomura2021Dirac-TypeSpectroscopy}, generated by VMC on an ansatz combining many-variable Gutzwiller-projected spinon-mean-field wave functions~\cite{mVMC2019} and restricted Boltzmann machines (RBMs). 
We achieve the most marked improvements in energy (comparable to the Hamiltonian gap) for the largest, $16\times16$ lattices:
While for small systems, a (small) linear combination of projected Slater determinants is an excellent approximation of spin-liquid and VBS ground states, this approximation deteriorates for larger lattices, beyond the expressivity of the RBM correction factors employed by Ref.~\cite{Nomura2021Dirac-TypeSpectroscopy}.
By contrast, the deep networks used by us are not subject to such a priori limitations.
Furthermore, the computational cost of our simulations are significantly smaller: for the $16\times16$ lattice in particular, we require approx.\ 300 GPU hours, compared to the approx.\ 60\,000 CPU hours of Ref.~\cite{Nomura2021Dirac-TypeSpectroscopy}.

To further verify the quality of our converged wave functions, we computed their total spin $\langle S^2\rangle$, expected to be exactly zero for an antiferromagnetic ground state. While we cannot project the wave function on the $S=0$ sector explicitly, we find $\langle S^2\rangle < 0.3$ even for the $16\times16$ lattice, a substantial improvement over $\langle S^2\rangle\approx0.6$ for $L=10$ reported earlier for various shallower NQS architectures~\cite{Choo2019StudyStates,Szabo2020NeuralProblem}.
$\langle S^2\rangle$ is consistently higher for larger system sizes and values of $J_2$, reflecting the increasing difficulty of obtaining accurate wave functions.
The spin structure factors $\langle \vec s(-\vec q) \cdot \vec s(\vec q)\rangle$ and $\langle s^z(-\vec q) s^z(\vec q)\rangle$ showed similarly good agreement for all wave vectors $\vec q$, showing that our wave functions do not strongly break SU(2) spin rotation symmetry, a common issue for earlier ansätze~\cite{Zhang2022HamiltonianReconstruction}.
Furthermore, we demonstrate in \cref{appendix: observables} that our method overcomes the ``sign problem'' of neural quantum states~\cite{Westerhout2020,Szabo2020NeuralProblem}, which has hampered the performance of such ansätze as recurrent neural networks for frustrated magnets~\cite{Hibat-Allah2020RecurrentWavefunctions}.

\subsection{Correlation functions}

\begin{figure}
    \centering
    \includegraphics{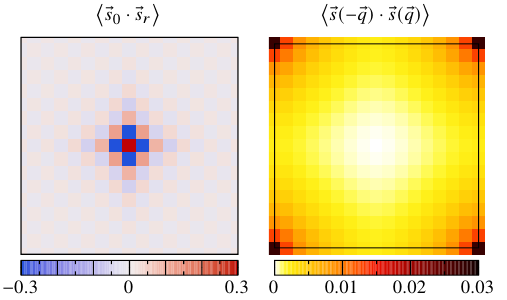}
    \caption{Spin correlation function for the square lattice $J_1-J_2$ model on a $16\times16$ lattice at $J_2=0.5$ in real space (left) and in the Brillouin zone (right). The true value of $\langle \vec s_0 \cdot \vec s_0 \rangle=3/4$ is clipped for visibility.}
    \label{fig: square spin correlation}
\end{figure}

We computed the real-space spin correlation function $\langle \vec s_0\cdot \vec s_r\rangle$ and its Fourier transform,
\begin{equation}
    S(\vec q) = \left\langle \vec s(\vec q) \cdot \vec s(-\vec q)\right\rangle = \frac1N \sum_{r} \langle \vec s_0 \cdot \vec s_{r}\rangle e^{-i\vec q\cdot \vec r},
    \label{eq: spin structure factor}
\end{equation}
and plotted them for the $16\times16$ cluster at $J_2=0.5$ in Fig.~\ref{fig: square spin correlation}. 
As expected for a spin-liquid state, the pronounced antiferromagnetic correlations between nearby sites decay rapidly.
Likewise, there are no pronounced Bragg peaks in reciprocal space, only a diffuse pattern with a smooth maximum at the $M$ point.
The picture for $J_2=0.55$ is very similar, with a somewhat faster decay of the real-space correlator and a less pronounced maximum at the $M$ point.

\begin{figure}
    \centering
    \includegraphics{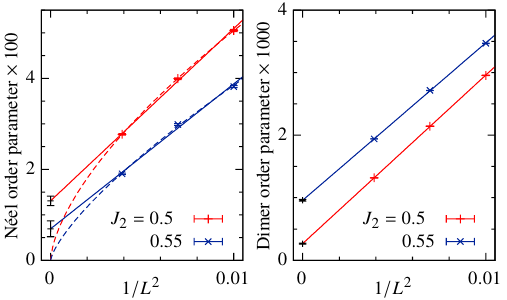}
    \caption{Néel order parameter $S(\pi,\pi)$ (left) and columnar dimer order parameter $\mathcal D$ (right) as a function of system size for the square-lattice $J_1-J_2$ model. Solid and dashed lines show scaling fits consistent with ordered~\eqref{eq: scaling ordered} and disordered~\eqref{eq: scaling power law} ground states, respectively.}
    \label{fig: square order parameter}
\end{figure}

To check for any residual spin ordering, we plotted the Néel order parameter $S(\pi,\pi)$ as a function of system size in Fig.~\ref{fig: square order parameter}. In an ordered phase, this is expected to scale as~\cite{Astrakhantsev2021Pyrochlore}
\begin{equation}
    S(L) \simeq S_0 + \alpha/L^2
    \label{eq: scaling ordered}
\end{equation}
The data do not fit this relationship very well (compared to the statistical error bars), and the error of the extrapolated order parameter is far larger than that of each data point.
By contrast, we expect
\begin{equation}
    S(L) \propto L^{-z}
    \label{eq: scaling power law}
\end{equation}
in a disordered phase; Ref.~\cite{Nomura2021Dirac-TypeSpectroscopy} estimated $z\approx1.5$ for the Néel order parameter. Our data points fit much better to such scaling too, indicating the lack of spin ordering at both $J_2=0.5$ and 0.55.

\begin{figure}
    \centering
    \includegraphics{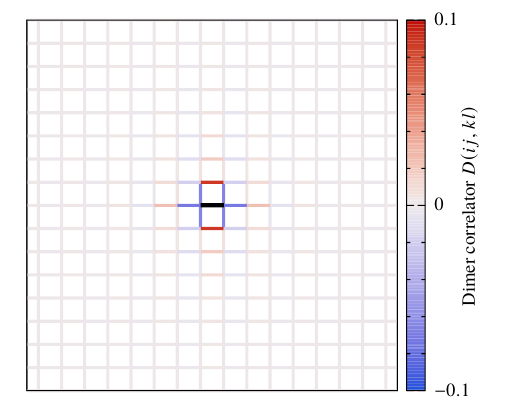}
    \caption{Real-space dimer correlation function~\eqref{eq: dimer correlator} for the square-lattice $J_1-J_2$ model on a $16\times16$ lattice at $J_2=0.5$ in real space. The reference bond $ij$ is shown in black; the true value of $D(ij,ij)= 0.2466(2)$ is clipped for visibility. The picture for $J_2=0.55$ is very similar; the dimer order parameter is too small to be directly visible.}
    \label{fig: square dimer correlation}
\end{figure}

We also computed the connected dimer correlation function
\begin{equation}
    D(ij,kl) = \big\langle (\vec s_i\cdot \vec s_j) (\vec s_k\cdot \vec s_l) \big\rangle - \langle \vec s_i\cdot \vec s_j \rangle\langle\vec s_k\cdot \vec s_l \rangle.
    \label{eq: dimer correlator}
\end{equation}
As $D$ can be rewritten as the covariance of local estimators of bond operators acting only on sites $i,j$ and $k,l$, respectively (\cref{appendix: observables}), we could compute it for all pairs of nearest-neighbour sites at modest computational cost, allowing for very accurate measurements.
These correlators are plotted for $J_2=0.5$ in Fig.~\ref{fig: square dimer correlation}: they show a short-range columnar pattern, consistent with Ref.~\cite{Ferrari2020GaplessExcitations}, but decaying extremely quickly with the separation of dimers. 

Even though we expect dimer ordering at $J_2=0.55$, the plot of $D(ij,kl)$ is almost identical. To probe ordering more carefully, we define the dimer order parameter
\begin{equation}
    \mathcal D = \frac1{2N^2}\sum_\mu \sum_{a,b\in \mu} D(a,b) e^{-i\vec q_\mu \cdot(\vec r_b-\vec r_a)},
\end{equation}
where $\mu=x,y$ are the possible orientations of nearest-neighbour bonds,
$a,b$ are bonds of the same orientation,
$\vec q_x = (\pi,0)$ and $\vec q_y=(0,\pi)$ are the columnar ordering wave vectors,
and $\vec r_{a,b}$ are a reference point on each bond (e.g., their midpoints). 
$\mathcal D$ is plotted in Fig.~\ref{fig: square order parameter}; it is quite small for every system we probed, consistent with the rapid decay of $D$ in real space.
For both $J_2=0.5$ and 0.55, $\mathcal D$ fits the scaling form~\eqref{eq: scaling ordered}.
The extrapolated order parameter is very close to zero for $J_2=0.5$; it may well be eliminated by fitting to~\eqref{eq: scaling power law} with $z\approx 2$, consistent with the findings of Ref.~\cite{Nomura2021Dirac-TypeSpectroscopy}. 
By contrast, the extrapolated $\mathcal D$ for $J_2=0.55$ is similar in magnitude to the finite-size measurements, which do not reasonably fit a power law, indicating dimer order.

\subsection{Excited states}
\label{sec: square excited}

We further benchmarked our method by obtaining variational energies in the symmetry sectors corresponding to the lowest-lying excitations of the two ordered phases bordering the Dirac spin-liquid phase.
In the Néel state, these come from a tower of states corresponding to the gapless magnon mode at the $M$ point~\cite{Anderson1952TOS}:
the lowest entry of this tower is a triplet transforming according to the $M.A_1$ irrep of the space group.
In the columnar valence-bond-solid proposed by Ref.~\cite{Ferrari2020GaplessExcitations}, breaking translational symmetry yields four degenerate singlet ground states, including a pair transforming according to the $X.B_2$ irrep, which become the lowest-lying excited states in a finite-size system.

\begin{table}[]
    \centering
    \setlength{\tabcolsep}{0.4em}
    \begin{tabular}{cccccc} \hline\hline
        \multirow{2}{*}{$J_2/J_1$} & \multirow{2}{*}{System size} & \multicolumn{2}{c}{Our work} & \multicolumn{2}{c}{Ref.~\cite{Nomura2021Dirac-TypeSpectroscopy}} \\
        && $M.A_1^-$ & $X.B_2^+$ & $M^-$ & $X^+$\\\hline
        \multirow{3}{*}{0.5} & 
        $10\times10$ & 0.2284(10) & 0.2755(11) & --- & ---\\
        &$12\times12$ & 0.1820(17) & 0.2460(16) & 0.1659(9) & 0.2282(4)\\
        &$16\times16$ & 0.1050(17) & 0.1867(21) & 0.0944(5) & 0.1788(5) \\\hline
        \multirow{3}{*}{0.55} & 
        $10\times10$ & 0.2838(14) & 0.2160(15) & --- & ---\\
        &$12\times12$ & 0.2170(20) & 0.1522(19) & 0.2202(4) & 0.1371(4)\\
        &$16\times16$ & 0.2004(27) & 0.1477(27) & 0.1620(8) & 0.0805(8) \\\hline\hline
    \end{tabular}
    \caption{Energy gaps (in units of $J_1$) to the optimised variational energies in several symmetry sectors for the square-lattice $J_1-J_2$ Heisenberg model, compared to the best known figures in the literature~\cite{Nomura2021Dirac-TypeSpectroscopy}. Most of our gaps are marginally (by up to 10\%) larger, owing to the lower ground-state energies we attain.}
    \label{tab: square excited}
\end{table}

We trained networks of the same structure as used for the ground-state estimates using the transfer learning procedure outlined in Sec.~\ref{sec: training protocol}.
Our best estimates for the gaps between the excited and ground-state energies in these two symmetry sectors are given in Table~\ref{tab: square excited}. 
These gaps all match the corresponding results of Ref.~\cite{Nomura2021Dirac-TypeSpectroscopy} within 10\%, suggesting that the necessarily more complex sign structure of the excited states poses little additional difficulty for the GCNN.

\section{Triangular-lattice $J_1-J_2$ model}

The phase diagram of the triangular-lattice Heisenberg antiferromagnet is similar to the square-lattice case:
the small and large $J_2$ limits show three-sublattice $120^\circ$ and stripy orders, respectively, with a quantum paramagnetic phase between the two orders near the point of maximal frustration, $J_2=J_1/8$~\cite{Iqbal2016TriangularJ1J2,Hu2019DiracAntiferromagnet,Wietek2017TriangularHeisenbergCSL}.
However, the triangular lattice is not bipartite: the resulting geometrical frustration heavily suppresses the $120^\circ$ order parameter even at $J_2=0$ and gives rise to dynamics not captured accurately by linear spin-wave theory~\cite{Ghioldi2018Triangular,Macdougal2020TriangularHAF}.
For the same reason, reliable numerical simulation is much more challenging than in the square-lattice case, and indeed, the extent and nature of the paramagnetic phase remains uncertain, with viable proposals of both spin-liquid and VBS ground states~\cite{Wietek2017TriangularHeisenbergCSL}.

We benchmarked our GCNN wave functions at the nearest-neighbour point $J_2=0$ and at the point of maximal frustration, $J_2=J_1/8$, by computing ground-state energies and correlation functions, as well as excited state gaps.
For the 36-site simulations, as well as the 108-site simulation at $J_2 = 0$, the plain GCNN introduced in Sec.~\ref{sec: architecture} was able to find accurate ground and excited states.
We used 4096 Monte Carlo samples and GCNNs with 4 (8) layers of 6 feature maps each for the 36- (108-) site simulations. We trained the 36-site models for 12 hours  (approx.\ 1500 steps) and the 108-site model for 4 days (approx.\ 2000 steps).

For the more challenging $J_2=J_1/8$ model, we found that even deep GCNNs struggled to resolve the ground states of the 108- and 144-site clusters with sufficient accuracy, and their training became unstable with increasing depth.
To remedy this, we used the residual GCNN architecture shown in Sec.~\ref{sec: residual gcnn}, with 12 residual blocks of six feature maps.
Instead of the standard SR step~\eqref{eq: basic SR}, we trained these very deep networks using a modified algorithm, dubbed LayerSR, discussed in detail in Appendix~\ref{appendix: layerSR}.
Since this algorithm only requires constructing only a small portion of the quantum geometric tensor at a time, we were able to use more Monte Carlo samples, which considerably improved the converged wave functions.
In particular, we used 4096 samples until the energy plateaued, which took 4 days (approx.\ 600 steps),
and then 16384 samples for an additional 8 days (approx.\ 300 steps).
After training, we further improved the variational energies by applying a Lanczos step as described in Appendix~\ref{appendix: Lanczos Step}.

\subsection{Ground-state energies}

\begin{table*}[]
    \centering
    \setlength{\tabcolsep}{0.6em}
    \begin{tabular}{cccccccc}\hline\hline
        $J_2/J_1$ & $N$ & GCNN & GCNN+LS & Exact/Interpolated~\cite{Bernu1994Triangular,Iqbal2016TriangularJ1J2} &  Graph NN~\cite{kochkov2021learning} & Gutzwiller\,+\,LS~\cite{Iqbal2016TriangularJ1J2} & $\langle S^2\rangle$ \\\hline
        \multirow{2}{*}{0} & 36 & $-0.560313(3)$ & --- & $-0.5603734$ & --- & --- & 0.0022(5) \\
        & 108 & $-0.55315(3)\ph$ & --- & --- & $-0.5519(4)$ & --- & 0.131(4)\ph \\\hline
        \multirow{3}{*}{$1/8$} & 36 & $-0.515386(9) $ & --- & $-0.515564\ph$ & $-0.5131(8)$ & $-0.512503(3)$ & 0.0067(8) \\
        & 108 & $-0.51175(7)\ph$ & $-0.51268(9)$ &  $-0.51297\phs$ & $-0.5069(8)$ & --- & 0.208(7)\ph \\ 
        & 144 & $-0.51101(6)\ph$ & $-0.51218(9)$ & $-0.51275\phs$ & --- & $-0.510558(5)$ & --- \\ 
        \hline\hline
    \end{tabular}
    \caption{Optimised ground-state energies (in units of $J_1/\mathrm{spin}$) and estimates of the total spin $\langle S^2\rangle$ for the $J_1-J_2$ triangular-lattice Heisenberg model, compared to exact and variational benchmarks in the literature. The ``exact/interpolated'' energies are obtained from the thermodynamic limit estimated in Ref.~\cite{Iqbal2016TriangularJ1J2} and the exact 36-site energy~\cite{Bernu1994Triangular}, assuming that finite-size effects scale as $1/L^3$~\cite{kochkov2021learning}. }
    \label{tab: triangular energy}
\end{table*}

Our variational energies are summarised in Table~\ref{tab: triangular energy}, together with the best variational benchmarks in the literature~\cite{Bernu1994Triangular,Iqbal2016TriangularJ1J2,kochkov2021learning}.
On the 36-site cluster, GCNNs achieve energies extremely close to exact-diagonalisation results for both the nearest-neighbour and maximally frustrated models within a few GPU-hours.
On the 108-site cluster too, GCNNs achieve state-of-the-art variational accuracy. GCNNs outperform other state-of-the-art neural-network benchmarks~\cite{kochkov2021learning} by a wide margin, especially at the highly frustrated point $J_2=J_1/8$. 
Our results also surpass those from Gutzwiller-projected fermionic mean-field ansätze with additional variational Lanczos steps~\cite{Iqbal2016TriangularJ1J2}.
Upon including one Lanczos step, we far outperform all other variational approaches and reach relative errors of about $10^{-3}$ compared to ground-state energy estimates extrapolated from exact diagonalisation.
We find that we need at least this level of precision in order to resolve excited-state gaps.             

\subsection{Correlation functions}

\begin{figure}
    \centering
    \includegraphics{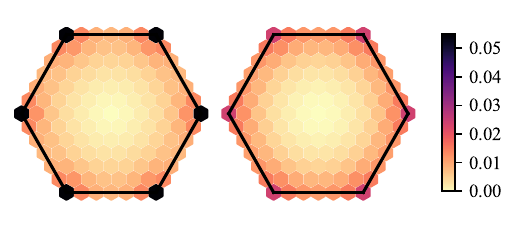}
    \caption{Spin structure factor in the Brillouin zone for the 108-site triangular cluster at $J_2 = 0$ (left) and $J_2 = J_1/8$ (right).}
    \label{fig: structure108}
\end{figure}

We computed the reciprocal-space spin structure factors $S(q)$
for the converged 108-site GCNN wave functions and plotted them in Fig.~\ref{fig: structure108}.
At $J_2=0$, we see strong Bragg peaks at the $K$ points, consistent with $120^\circ$ order;
their relatively low intensity and the strong diffuse component is compatible with a small ordered moment suppressed by frustration~\cite{Bernu1994Triangular,Singh1992TriangularKagome}.
At the point of maximal frustration, we find no distinct Bragg peak at either the $K$ points or the $M$ points expected for the large-$J_2$ stripy phase. 
Instead, we see a broad continuum around the edges of the Brillouin zone, consistent with an intermediate quantum paramagnetic phase~\cite{Iqbal2016TriangularJ1J2,Wietek2017TriangularHeisenbergCSL,Hu2019DiracAntiferromagnet}.

\begin{figure} 
    \centering
    \includegraphics{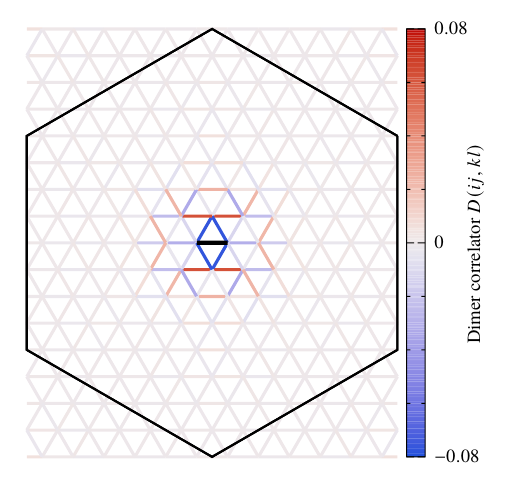}
    \caption{Real-space dimer correlation function~\eqref{eq: dimer correlator} for the triangular-lattice $J_1-J_2$ model on a 144-site lattice at $J_2=0.5$ in real space. The reference bond $ij$ is shown in black; the true value of $D(ij,ij)= 0.2445(8)$ is clipped for visibility.}
    \label{fig: dimer108}
\end{figure}

To gain further insight into this intermediate phase, we computed the connected dimer-dimer correlators $D(ij,kl)$
for all relative positions of the nearest-neighbour pairs $ij$ and $kl$. These are plotted in Fig.~\ref{fig: dimer108}: dimer correlations appear to decay to very small values over a correlation length of about three unit cells. 
While simulations on larger system sizes would be needed to exclude the possibility of a small residual VBS order parameter, these results are most compatible with a spin-liquid state.

\subsection{Excited states}

In addition to the ground state, we also estimated the energy of low-lying excited states in nontrivial symmetry sectors using the transfer learning procedure described in Sec.~\ref{sec: training protocol}.
The tower of states for the $120^\circ$ order is generated by gapless magnon modes at the $K$ points: the lowest-lying entries are triplets transforming under the $K.A_1$ and $\Gamma.B_1$ irreps of the space group~\cite{Wietek2017TOS}.
For the stripy phase at large $J_2$, the lowest lying triplet excitation transforms as $M.A_1$; in addition, the spontaneous breaking of point-group symmetry gives rise to a $\Gamma.E_2$ singlet.
The proposed intermediate Dirac spin liquid phase has a plethora of gapless excitations that can be captured either as pairs of gapless spinons or monopoles of the $U(1)$ gauge field~\cite{Song2019Monopoles,Song2020Monopoles}; all states mentioned above lie in this manifold. 

Our numerical results are summarised in Table~\ref{tab: triangular gaps}. 
On 36-site clusters, we recover the exact gaps~\cite{Wietek2017TriangularHeisenbergCSL,Wietek2017TOS} to a good approximation.
For $J_2=0$, the gap estimate of the $K.A_1^-$ state, corresponding to $120^\circ$ order, is seen to decrease, while the others remain approximately the same, as expected for the ordered phase.

In the paramagnetic phase, we find that when we supplement our wavefunctions with a Lanczos step, the gaps for all three irreps decrease going from 36 to 108 sites, indicating that they may be gapless modes.
However, they do not appear to scale as $N^{-1/2}$, as would be expected for a Dirac spin liquid~\cite{wietek2023quantum}.
This may be partially related to differences between the 36- and 108-site geometries,
but we believe it is mostly due to the increased difficulty to obtain accurate variational energies on larger clusters.
Indeed, we find that all of the gaps of all three irreps decrease when a Lanczos step is applied, indicating that the variational excited states are less accurate than the ground state.

\begin{table}[]
    \centering
    \setlength{\tabcolsep}{0.5em}
    \begin{tabular}{ccccc} \hline\hline
        $J_2/J_1$ & System size & $K.A_1^-$ & $M.A_1^-$ & $\Gamma.E_2^+$ \\\hline
        \multirow{2}{*}{0} & 36 & 0.3499(5)\ph & 0.9064(7)\ph & 0.9040(13)  \\
        & 108 & 0.198(6)\phs & 0.780(7)\phs & 1.039(11)\ph \\\hline
        \multirow{4}{*}{$1/8$} & 36 & 0.4865(12) & 0.5906(16) & 0.2206(7)\ph \\
        & 108 & 0.396(12)\ph & 0.591(13)\ph & 0.242(11)\ph \\
        & 108 + LS & 0.350(15)\ph & 0.414(16)\ph & 0.176(20)\ph \\\hline\hline
    \end{tabular}
    \caption{Energy gaps (in units of $J_1$) to the optimised variational energies in several symmetry sectors for the triangular-lattice $J_1-J_2$ Heisenberg model. For $J_2 = {J_1}/{8}$, we list the gap both before and after applying a Lanczos step.}
    \label{tab: triangular gaps}
\end{table}

\section{Conclusion}

In summary, we demonstrated the power of deep neural networks to represent highly entangled many-body wave functions.
In particular, we used group convolutional neural networks (GCNNs) to study the $J_1-J_2$ Heisenberg model on the square and triangular lattices, both of which are thought to host a quantum spin-liquid phase.
We demonstrated that GCNNs achieve competitive variational energies in the quantum paramagnetic phases of the square lattice with a relativity modest computational effort.
For the more difficult triangular lattice model, we showed that GCNNs with residual layers could improve upon previous state-of-the-art results by almost an order of magnitude, especially when supplemented with a Lanczos step.
These results highlight deep neural networks, and GCNNs in particular, as promising ansätze to study challenging quantum many-body problems.  

While we were able to consistently achieve excellent ground-state variational energies, doing so for excited states proved significantly harder.
Projecting on nontrivial irreps of the space group often leads to an unstable training trajectory, as well as excessive variational gap estimates. In future work, we plan to understand the origin of this issue and propose ways to remedy it in order to gain reliable access to the low-lying spectrum, a key diagnostic for establishing phase diagrams.

\textit{Note added.---}While revising this manuscript, we became aware of Ref.~\cite{Reh2023NQSdesign}, which systematically studies NQS design choices, including the role of the sign structure for frustrated systems. Their numerical results support the arguments leading to our choice of architecture in Sec.~\ref{sec: architecture}.

\begin{acknowledgements}
    We thank Marin Bukov and Alexander Wietek for helpful discussions.
    We are especially grateful to Yusuke Nomura for sharing detailed numerical results from Ref.~\cite{Nomura2021Dirac-TypeSpectroscopy} with us.  
    Simulations were performed using the NetKet~\cite{Vicentini2022NetKetSystems} library.
    All heat maps use perceptionally uniform colour maps developed in Ref.~\cite{Kovesi2015GoodThem}.
    Computing resources were provided by STFC Scientific Computing Department’s SCARF cluster and  STFC Cloud service, as well as the Texas Advanced Computing Center (TACC) at the University of Texas at Austin.  
    A.~Sz.\ gratefully acknowledges the ISIS Neutron and Muon Source and the Oxford--ShanghaiTech collaboration for support of the Keeley--Rutherford fellowship at Wadham College, Oxford. 
    C.~R.\ acknowledges support from the University of Michigan Grant No. 2635103712 and Cornell University Grant No. 2635160012. 
    For the purpose of open access, the authors have applied a Creative Commons Attribution (CC-BY) licence to any author accepted manuscript version arising.
\end{acknowledgements}

\appendix

\section{Gradients of the computational-basis entropy}
\label{sec: gradients}

Consider a wave function ansatz $\psi(\vec\sigma; \theta)$ with real parameters $\theta$. For brevity, we introduce $p(\vec\sigma) \equiv |\psi(\vec\sigma; \theta)|^2$, $P=\sum_\sigma p(\vec\sigma)$, and $q(\vec\sigma) = p(\vec\sigma)/P$. Now, the derivative of~\eqref{eq: entropy} with respect to $\theta$ is
\begin{align}
    \partial_\theta \mathcal{S} &= -\sum_{\vec\sigma} \left[\partial_\theta q(\vec\sigma) \log q(\vec\sigma) + q(\vec\sigma) \frac{\partial_\theta q(\vec\sigma)}{q(\vec\sigma)}\right] \nonumber\\
    &= -\sum_{\vec\sigma} \partial_\theta q(\vec\sigma) \log q(\vec\sigma) - \partial_\theta\Big[ \sum_{\vec\sigma} q(\vec\sigma) \Big] \nonumber\\
    &= -\sum_{\vec\sigma} \partial_\theta q(\vec\sigma) \log p(\vec\sigma) + \partial_\theta\Big[ \sum_{\vec\sigma} q(\vec\sigma) \Big] \log P \nonumber\\
    &= -\sum_{\vec\sigma} \left[\frac{\partial_\theta p(\vec\sigma)}P - \frac{p(\vec\sigma)}P\frac{\partial_\theta P}{P} \right] \log p(\vec\sigma) \nonumber\\
    &= -\langle \log p\ \partial_\theta(\log p)\rangle + \langle \log p\rangle\langle\partial_\theta(\log p)\rangle,
    \label{eq: entropy gradient}
\end{align}
where the final expectation values are taken with respect to the Born distribution $q(\vec\sigma)$.
We repeatedly use the fact that $\sum_{\vec\sigma} q(\vec\sigma) = 1$, so its derivative vanishes.
\cref{eq: entropy gradient} gives a Monte Carlo estimator for $\partial_\theta \mathcal{S}$ that can be used for unnormalised ansätze.
Furthermore, as $\log p = 2\Re \log \psi$, we can incorporate~\eqref{eq: entropy gradient} in the usual Monte Carlo estimator of the energy gradient~\cite{becca_sorella_2017} at minimal computational overhead by adding $T\log p$ to the local energy $\langle \vec\sigma|H|\psi\rangle/\langle \vec\sigma|\psi\rangle$.

\section{Computing spin and dimer correlation functions}
\label{appendix: observables}

The expectation value of any Hermitian operator $\hat O$ with respect to a variational wave function can be evaluated as the Monte Carlo average of the local estimator
\begin{equation}
    O_\mathrm{loc}(\vec\sigma) = \frac{\langle \vec\sigma | \hat O | \psi\rangle}{\langle \vec\sigma | \psi\rangle}
    \label{eq: local estimator}
\end{equation}
with respect to the Born distribution $q(\vec\sigma)$ introduced above~\cite{becca_sorella_2017}.
Furthermore, the variance of the local estimators converges to the quantum mechanical variance $\var\hat O = \langle \hat O^2\rangle -\langle \hat O\rangle^2$.
We used this directly for $\hat O = \vec s_i \cdot \vec s_j$ to estimate real-space spin correlators.
To limit the effect of not fully translationally invariant sampling on the results, we computed $\langle \vec s_i\cdot\vec s_j\rangle$ for all pairs of spins, at the expense of using fewer independent Monte Carlo samples.

To ensure that our wave functions do not break spin-rotation symmetry, we also computed $\langle s_i^z s_j^z\rangle$ for the square-lattice ground state, expected to be one third of the corresponding $\langle \vec s_i\cdot\vec s_j\rangle$.
Since these operators are diagonal in the computational basis, they can be computed as the covariance of the appropriate $\sigma^z$ in the sampled bit strings $\vec\sigma$.
Not having to compute local estimators using~\eqref{eq: local estimator} allows us to use significantly more samples at the same computational cost.
Their variance can be obtained by noting that $(s_i^zs_j^z)^2$ is a constant (in the spin-operator normalisation, $1/16$).

Making use of the identity
\begin{align}
    \langle \hat A \hat B \rangle &= \frac1{\langle \psi|\psi\rangle}\sum_{\vec \sigma} \langle \psi | \hat A|\vec\sigma\rangle\langle \vec\sigma | \hat B|\psi\rangle \nonumber\\
    &= \sum_{\vec\sigma} q(\vec\sigma) A_\mathrm{loc}^*(\vec\sigma) B_\mathrm{loc}(\vec\sigma),
\end{align}
dimer correlators~\eqref{eq: dimer correlator} were computed as the covariance of the local estimators of $\vec s_i\cdot\vec s_j$ and $\vec s_k\cdot\vec s_l$. The advantage of this method over computing $\big\langle (\vec s_i\cdot \vec s_j) (\vec s_k\cdot \vec s_l) \big\rangle$ directly using~\eqref{eq: local estimator} is that computing the local estimators for each nearest-neighbour bond operator $O(N_\mathrm{sample} N)$ yields all $O(N^2)$ dimer correlators, allowing us to use substantially more independent Monte Carlo samples. 
(Computing the covariance takes negligible time compared to the wave function evaluations needed for sampling and computing the local estimators.)
However, this method does not allow us to directly obtain the variance of $D(ij,kl)$; instead, we estimated its error from the variance of translationally equivalent estimates.

\begin{figure}
    \centering
    \includegraphics{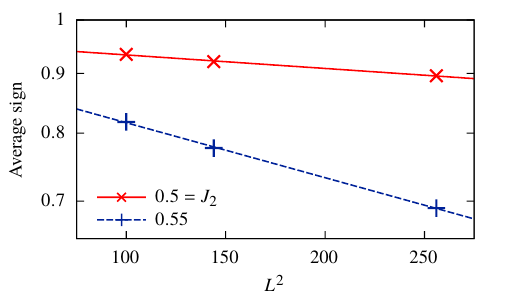}
    \caption{Average sign of the best converged wave function after Marshall transformation. For both values of $J_2$, they are very closely exponential in system size (fitted lines). The error bars are smaller than the symbols.}
    \label{fig: average sign}
\end{figure}

On the square lattice, the sign structure of the $J_2=0$ ground state is governed by the Marshall sign rule~\cite{Marshall1955Antiferromagnetism};
the frustrated $J_1-J_2$ model, by contrast, has a nontrivial sign structure even after applying the Marshall transformation.
Monte Carlo estimates of the Marshall-adjusted average sign for our best converged wave functions are plotted in Fig.~\ref{fig: average sign}:
they decay exponentially in the number of sites, as expected for a sign-problematic Hamiltonian~\cite{Troyer2005ComputationalSimulations}.
The decay is much faster at  $J_2=0.55$, consistent with the higher degree of frustration.

For the square-lattice wave functions, we used $2^{11}$ Monte Carlo samples to compute $\langle \vec s_i\cdot\vec s_j\rangle$ and the same $2^{18}$ samples to obtain $\langle s_i^z s_j^z\rangle$, $D(ij,kl)$, and the average sign.
For the 36- (108-) site triangular lattice, we used $2^{17}$ ($2^{14}$) samples for both spin and dimer correlators.

\section{LayerSR algorithm}
\label{appendix: layerSR}

To train the residual GCNN ansätze, we use an alternative to the standard stochastic reconfiguration~\eqref{eq: basic SR}.
As shown in Ref.~\cite{chen2023efficient}, if the number of samples is less than the number of neural-network parameters, Eq.~\eqref{eq: basic SR} is equivalent to
\begin{equation}
    \overline{O}\, \delta\vec\theta = -\eta \overline{E}_\mathrm{loc},
    \label{eq: SR alternative}
\end{equation}
where $E_\mathrm{loc}$ is the local estimator~\eqref{eq: local estimator} of the Hamiltonian,
$O = [\partial\log(\psi_i)/\partial\theta_j]_{ij}$ is the Jacobian of $\log\psi$ with respect to the parameters $\theta$, and the overline stands for subtracting the mean across samples:
\begin{align}
    \overline{O}_{ij} &= O_{ij} - \langle O_j\rangle; &
    \overline{E}_{i,\mathrm{loc}} &= {E}_{i,\mathrm{loc}} - \langle {E}_{\mathrm{loc}}\rangle.
\end{align}
If the number of samples is less than the number of neural-network parameters, Eq.~\eqref{eq: SR alternative} is also underdetermined, so we are free to constrain the parameter updates further.

For LayerSR, we require that the updates $\delta\vec\theta_\mathrm{layer}$ of the parameters of each layer in the network satisfy
\begin{equation}
    \overline{O}_\mathrm{layer}\, \delta\vec\theta_\mathrm{layer} = -\frac{\eta \overline{E}_\mathrm{loc}}{n_\mathrm{layer}},
    \label{eq: SR layer}
\end{equation}
where $n_\mathrm{layer}$ is the number of layers.
Summing~\eqref{eq: SR layer} for all layers recovers~\eqref{eq: SR alternative},
so as long as there are fewer samples than parameters in each layer [i.e., Eq.~\eqref{eq: SR layer} is underdetermined too], the solutions of the former give a valid solution of the latter too. 
Eq.~\eqref{eq: SR layer} can be recast in the standard SR form~\eqref{eq: basic SR}, which we solved using the regularisation described in the main text.

Computationally, LayerSR improves on the implementations of standard SR available in NetKet~\cite{Vicentini2022NetKetSystems}, which either recompute the Jacobian $O$ on the fly many times (and are thus impractically slow for a deep network) or store it in full (causing severe memory limitations on GPUs).
LayerSR only requires storing the Jacobian of a single layer $O^{\textrm{layer}}$ in memory, which results in a memory cost reduction proportional to the number of layers at the modest cost of performing the backpropagation needed to obtain $O$ once for each layer.
This allows us to use more samples on the single GPU, leading to significantly better converged energies.
We also found that LayerSR tends to improve the convergence of the GCNN with residual layers during the early stages of training. 

\section{Implementation of the Lanczos step}
\label{appendix: Lanczos Step}

We implement one Lanczos step for the optimised GCNN wave functions $|\psi_0\rangle$; 
that is, we find the wave function with minimum variational energy in the two-dimensional Krylov space spanned by $|\psi_0\rangle$ and the orthogonal component of $H|\psi_0\rangle$,
\begin{equation}
|\psi_1\rangle =  \frac{H-\langle H \rangle}{\sigma} |\psi_0 \rangle,
\end{equation}
which we normalised with the variance $\sigma^2 = \langle H^2\rangle - \langle H \rangle^2$.
We thus have to diagonalise the $2\times2$ matrix 
\begin{align}
    H_\mathrm{L} &= 
\begin{pmatrix}
\langle \psi_0 | H | \psi_0 \rangle & \langle \psi_0 | H | \psi_1 \rangle \\
\langle \psi_1 | H | \psi_0 \rangle & \langle \psi_1 | H | \psi_1 \rangle
\end{pmatrix} \nonumber\\
&= \langle H\rangle + 
\begin{pmatrix}
\langle \psi_0 | H-\langle H\rangle | \psi_0 \rangle & \langle \psi_0 | H-\langle H\rangle | \psi_1 \rangle \\
\langle \psi_1 | H-\langle H\rangle | \psi_0 \rangle & \langle \psi_1 | H-\langle H\rangle | \psi_1 \rangle
\end{pmatrix} \nonumber\\
&= \langle H\rangle + \sigma 
\begin{pmatrix}
0  & 1 \\
1 & \mu_3
\end{pmatrix},
\label{eq: Lanczos matrix}
\end{align}
where $\mu_3 = \langle (H - \langle H \rangle)^3 \rangle/\sigma^3$ is the standardised third moment of the Hamiltonian acting on $|\psi_0\rangle$.
The lower eigenvalue of~\eqref{eq: Lanczos matrix} is
\begin{equation}
    E_\mathrm{L} = \langle H\rangle + \sigma \left(\frac{\mu_3}2 - \sqrt{\frac{\mu_3^2}4 + 1}\right),
\end{equation}
which is the first-order Lanczos estimate of the ground-state energy.
We compute $\sigma$ and $\mu_3$ by finding the local estimators~\eqref{eq: local estimator} of the operators $H-\langle H\rangle$ and $(H-\langle H\rangle)^2$, labelled $\varepsilon_\mathrm{loc}$ and $\nu_\mathrm{loc}$, respectively, for $2^{14}$ Monte Carlo samples, from which we estimate
\begin{align}
    \sigma^2 &= \langle \nu_\mathrm{loc}\rangle; &
    \mu_3 &= \frac{\cov(\varepsilon_\mathrm{loc},\nu_\mathrm{loc})}{\sigma^3}.
\end{align} 

This procedure is slightly different from Refs.~\cite{chen2023systematic, gagliano1986correlation}, which use local estimators of $H$ and $H^2$ (without the mean subtraction) to compute estimates of $\langle H^n \rangle$.
This introduces an ambiguity in computing $\langle H^2\rangle$ from the local estimators of either $H$ or $H^2$:
We found that the combination of the two implied by subtracting the mean from our estimators significantly improves statistical accuracy.

\bibliography{references,equivariant,netket,extra}

\end{document}